# Three-Stage Synthesis of a Cobalt-Embedded Graphene-like Carbon Framework with Long-Range Atomic Order

G.G. Ryzhkova[1,2], I.Yu. Kurochkin[1], T.N. Rudneva[1], A.V. Zotov[1], A.V. Moiseenko[3], G.M. Zirnik[4], D.A. Vinnik[4], V.I. Korepanov[1]

[1]Institute of Microelectronics Technology RAS, 6 Academician Ossipyan str., 142432 Chernogolovka, Russia.
[2]Samara National Research University, 34, Moskovskoye shosse, 443086 Samara, Russia.
[3]Faculty of Biology, Lomonosov Moscow State University, Moscow 119991, Russia.
[4]Moscow Institute of Physics and Technology, 9, Institutskij lane, 141701 Dolgoprudny, Russia.
*Corresponding author E-mail: korepanov@iptm.ru*

**Abstract**

We report a three-stage synthesis of a hybrid metal-carbon 2D material, in which cobalt atoms are covalently embedded in the graphene-like carbon (GLC) matrix. The resulting material (CoGLC) exhibits a distinctive XRD pattern indicative of the ordered arrangement of cobalt atoms in the layers. Furthermore, we demonstrate the fabrication of surfactant-free conductive inks from CoGLC via electrochemical exfoliation, making it a promising candidate for applications in in flexible electronics, spintronics or electrocatalysis.

# INTRODUCTION

Graphene-like carbon (GLC) materials are pivotal in advancing modern microelectronics. This is primarily due to their great potential for various structural modifications [1]. They have promising applications in the creation of sensors [2], transistors [3], capacitors [4], [5], spintronic devices [6]. At the same time, amorphous structures based on 2D materials are no less interesting than crystalline ones [7], [8].

The most common method of modifying graphene involves changing the structure of the layers by replacing carbon atoms with heteroatoms and/or atoms of various metals [9], [10]. Functionalization can be achieved by various methods. The most widely used methods include substitution doping using wet chemistry methods, hydrothermal synthesis, thermal annealing with doping gases, arc discharge and chemical vapor deposition (CVD) [5], [11], [12]. Meanwhile, methods using precursors with the required chemical elements in graphene synthesis are considered more effective for obtaining uniformly doped layers (such as CVD or arc discharge) [13].

However, metal atoms are more often introduced by functionalizing graphene/graphene already doped with heteroatoms (impregnation, reduction, or microwave treatment) [14], [15]. An efficient way of bonding metal atoms in the structure of graphene sheets is the use of heteroatoms [16], [17]. Nitrogen- or boron-doped graphenes are most suitable for coordinating metals. This is due to the fact that nitrogen and boron atoms have similar sizes and numbers of valence electrons to carbon, which allows them to more easily incorporate into the graphene structure.

Introducing heteroatoms or metals into graphene inherently creates defects in the sp² carbon network. The main problems are related to the non-uniform distribution of incorporated atoms in the sheets. A typical problem is that metals fail to bond with lighter atoms, which often leads to their agglomeration or clustering in nanoparticles [18]. Therefore, synthesizing a uniformly metal-doped 2D material with well-defined properties remains a significant challenge.

In the context of synthesizing less disordered metal-doped graphene-like materials, porphyrinoid compounds look promising as precursors. Such compounds themselves are of high interest for wide range of applications. This is due to a specific chemical structure, which includes a wide π-conjugated system and the large coordination cavity. Such geometry allows them to coordinate a different metals in the macrocyclic core [19]. There are several works in the literature that have attempted to introduce structure elements of such type into graphene. For example, the work [16] demonstrates the binding of Cu, Ag, and Au using the carbaporphyrin fragment incorporated into a graphene-like structure. In the work [20], the authors obtained graphene nanoribbons doped with Ni atoms, which was achieved using Ni-porphyrin polymer. In the work [21], the graphene material modified with metal-free porphyrin fragments was synthesized.

As demonstrated in a number of studies, it is possible to obtain polymeric structures based on porphyrinoid [22]. Meanwhile, phthalocyanines and polyphthalocyanines can serve as more abundant source of nitrogen in the synthesis of such 2D materials. Moreover, the amount of metal in the polymer structure that is almost stoichiometric ($C_{20}N_4Me$) can be achieved [23].

Despite their theoretically excellent properties, these compounds are difficult to treat, since they are powders that are practically insoluble in organic solvents. As a result, it is challenging to obtain high-quality films with high crystallinity based on such materials [24]. Compared to highly ordered crystalline structures, the resistance of such materials in an amorphous state increases by orders of magnitude. For these reasons, polyphthalocyanines themselves seems to be limited in their applicability in microelectronics. In this work, we address these challenges by modifying the polymer structures. As shown in a number of studies [25], [26], [27], heat treatment of such compounds in an inert environment leads to the loss of the pristine structure of polyphthalocyanine and the formation of graphene-like sheets. The main risks associated with carbonization procedure are also related to the possibility of metal atoms detaching from polymer network with subsequent agglomeration of the their into nanoparticles [27].

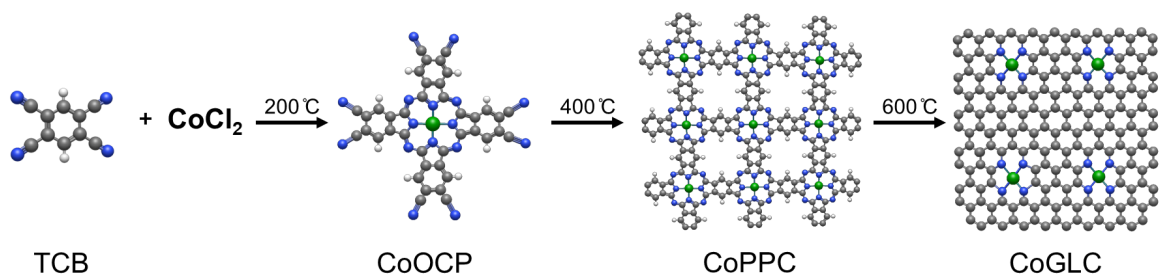

Figure 1. Reaction scheme of the three-stage synthesis to produce CoGLC

Thus, polyphthalocyanines remain the good candidate for the production of amorphous semiconductor 2D-materials with customized properties. Thus, they may be relevant and applicable in printed microelectronics [28] or areas of materials science where large active surface areas are important. In this article, we demonstrate the possibility of obtaining a graphene-like structure that is significantly doped with cobalt atoms and also demonstrates a surprising regularity in the distribution of metal atoms in nanosheets. The study presents the idea of synthesizing graphene-like materials doped with metal atoms, including the stage of direct synthesis from a metal-containing precursor (Fig. 1). One more objective of the present study was to solve the processability problem by making conductive inks from the exfoliated 2D material.

**EXPERIMENTAL DETAILS**

*Materials*. Commercial reagents were used for the synthesis: 1,2,4,5-tetracyanobenzene (TCB) (Sigma-Aldrich), cobalt chloride $CoCl_2·6H_2O$ (Merck), potassium bromide KBr 99% (JSC "Ural

Chemical Reagent Base", Russia), tetrabutylammonium perchlorate (TBA), $C_{16}H_{36}N^+ClO_4^-$ (98%, Macklin, China), argon (99.999%, Linde gas,Russia). Organic solvents: ethyl alcohol and acetonitrile were purchased from Chimmed (Russia), isopropanol and o-xylene from Ecos-1 (Russia), N-methyl-2-pyrrolidone (NMP) from Sisco Research Lab. (India).

*Cobalt graphene-like carbon (CoGLC) synthesis.* Cobalt chloride was dehydrated by heating to 120 ºC in a drying oven, during which it changed color from pink to lilac. Then TCB (50 mg), pre-dried cobalt chloride $CoCl_2$ (73 mg, which is a fourfold excess) and KBr (500 mg) were mixed ground in a mortar and sealed in an ampoule under vacuum. Each ampoule was loaded in a 12 caliber shotgun casing and plugged with a foil cap (Fig. S0). The ampoule was heated in an oven for 2 hours at a temperature of 200 ºC (the 1st stage, OCP), then the temperature was raised to 400 ºC and kept for another 2 hours (the 2nd stage, PPC). Finally, the ampoule with the mixture, without removing it from the oven, was further heated for another 2 hours at a temperature of 600 ºC (3rd stage, CoGLC. At this step, some of the ampoules might explode with a kaboom-like sound; typically, no one gets hurt). The ampoules from the three stages were treated separately in the following way: each ampoule was cooled and opened (this should be done carefully, since upon decompression a mild kaboom-like sound is emitted; typically, no one gets hurt). The reaction mixture was washed with hot deionized water and isopropanol, dried, ground in a mortar and again washed with water and isopropanol, then dried. The yield at stage 3 was 48 mg.

*Electrochemical exfoliation of CoGLC.* A two-electrode electrochemical cell was used to obtain 2D *CoGLC*. A platinum wire served as the anode, and *CoGLC* powder (45-50 mg) served as the cathode. Acetonitrile served as the solvent (40 ml). Exfoliation was performed in the presence of TBA as an exfoliating agent (0.05M). The operating voltage ranged from 15-30 V. The resulting material was isolated by centrifugation and washed with chloroform, ethanol, and deionized water to remove TBA traces and its reduction products.

*Ink preparation.* 15ml of NMP was added to the *CoGLC* powder from the previous step and sonicated for half an hour at 200 W while cooling on ice. The resulting mixture was centrifuged for 20 minutes at 1000 rpm to separate out any oversized particles. The suspension was then transferred to a separating funnel, and 8 ml of o-xylene and 12 ml of deionized water were added to form a Pickering emulsion. The middle fraction was separated and washed with alcohol. After drying, the polymer powder was mixed with a 0,6 ml monoethylene glycol. The resulting mixture was then mechanically homogenized until a homogeneous gel was formed.

*Raman spectroscopy.* Raman spectra were measured with a Bruker Senterra micro-Raman system under 532 nm excitation. Laser power was 0.1 mW at the sample point; acquisition time was 2x60s.

*IR spectroscopy.* Infrared absorption were measured at room temperature by FT-IR spectrometer Vertex 80V (Bruker Optics, Germany) in the range of 400-4000 $cm^{-1}$.. For each sample, 3 mg per 300 mg of KBr was taken to make a pellet.

*X-ray diffraction.*

Powder diffraction patterns were recorded with a Drawell DW-XRD-2700A diffractometer; CuKα line was used at 40kV and 30mA; scan speed was 1 °/min.

*Thermogravimetry-differential scanning calorimetry TG-DSC.*

The samples were analyzed using DTA/TGA on an STA 8000 synchronous thermal analyzer (Perkin-Elmer, USA) coupled to a Spectrum Two FTIR spectrometer (Perkin-Elmer). For CoOCP, the simultaneous analysis of volatile decomposition products was recorded in range of 20–800°C. The samples were placed in corundum crucibles; heating rate was 10°C/min or 15°C/min. Simultaneous analysis and recording of volatile thermal decomposition products were performed using the Pyris and TimeBase software packages (Perkin-Elmer).

*X-ray photo electron spectroscopy (XPS).* The chemical composition of the CoGLC inks was studied using X-ray photoelectron spectroscopy on an SPECS (Berlin, Germany) instrument with the FOIBOS-150 energy analyzer using Al-K-α excitation ($E_{ex}$ = 1486.61 eV). The spectra were calibrated to the ambient carbon C1s energy (285 eV) and analyzed with Casa XPS software (version 2.3.25PR1). The measurements were conducted with an anode voltage of 12.5 kV of the XR50 X-ray source and an emission current of 19.1 mA. The vacuum in the energy analyzer chamber of spectrometer's was maintained at $5 \times 10^{-9}$ Torr. Detailed recordings of the core level spectra were performed with a sweep step of 0.1 eV and an analyzer transmission energy (Epass) of 15 eV. The study utilized the Large Area lens mode. The lens mode of the energy analyzer used prevented foreign substances, such as the sample holder, from entering the analysis area. The ink sample was applied to gold-coated silicon substrate and dried in a laminar flow box on a furnace surface maintained at 180 °C. Afterwards, the substrate was fixed with a double-sided carbon tape to the holder's surface and loaded into the spectrometer. In calculating the elemental concentrations (at.%), the photoionization cross sections of core electrons from the SpecsLab library were taken into account. The uncertainty in determining the concentration is estimated to be ± 5% of the measured value.

*Transmission electron microscopy.* A 1 μL aliquot of the sample suspension deposited onto a carbon support film and subjected to vacuum drying at $10^{-5}$ mbar for 1 h. Imaging was performed on a JEM-2100 (JEOL, Japan) 200 kV $LaB_6$ TEM equipped with a DE-20 camera (Direct Electron, USA). Electron energy-loss spectra were obtained using a GIF Quantum spectrometer (Gatan, USA) with an energy dispersion of 0.25 eV per channel and a collection semi-angle of 0.5 mrad. Elemental maps were acquired in STEM mode with simultaneous HAADF imaging and spectrum imaging (one spectrum per pixel). To create a qualitative distribution map of cobalt, the Co $L_{2,3}$ edge was analyzed: a power-law background was fit over the 740–768 eV pre-edge region, and the signal was integrated from 778 to 828 eV under the $L_{2,3}$ edge. No multiple-scattering correction was applied, as the analysis was exploratory and qualitative. The data processing was performed using Digital Micrograph 2.3.2 software (Gatan, USA)

## RESULTS AND DISCUSSION
**Synthesis and thermal behavior**

The synthetic approach to make CoGLC starts with the polymerization of tetracyanobenzene (TCB). At low temperatures (about 200 °C, stage 1), the reaction of TCB with $CoCl_2$ yields an octacyanophthalocyanine (OCP), a single-ring substituted phthalocyanine. At higher temperatures, the reaction proceeds to a phthalocyanine polymer. The second stage of this reaction has a higher activation energy and requires at least 340 °C [29]; typically the reaction is performed at about 400 °C [30], [31]. At higher temperatures, metal-phthalocyanine polymers (MePPCs) can decompose, releasing hydrogen and nitrogen and forming layers of graphene-like material, and also yielding metal nanoparticles. The latter can be observed by XRD for the reaction temperature above 700 °C [26]. According to other data, metal reduction occurs at a temperature of 760 °C [32], or, as claimed in the work [33], at above 800 °C. The corresponding monomer, CoPC, decomposes in inert atmosphere at above 400 C [34]. Since the present work aimed at the synthesis of metal-linked graphene-like carbon, the reaction conditions should be chosen based on the thermal behavior of the reactions shown in Fig. 1.

TG/DSC analysis of CoOCP in an inert atmosphere revealed a multi-stage weight loss (Fig. S1). The first two stages (~260 °C and ~370 °C) are accompanied by release of gas products, containing the C≡N bonds, as confirmed by IR spectroscopy of the gaseous products (Fig. S2-S3). These products might include TCB, OCP, HCN and NCCN. A more significant decomposition step begins at 625 °C. Notably, an endothermic event at ~560 °C, immediately preceding this decomposition, is hypothesized to correspond to the reconstruction of the 2D polymer into a graphene-like structure. Based on this, we selected 600 °C as the optimal carbonization temperature to form the target CoGLC material while preserving the metal-carbon framework.

## XRD

The structural changes in the three stages of the synthesis were tracked by X-ray diffraction studies. The first-stage low-temperature product (CoOCP) shows a number of pronounced sharp peaks typical for low-molecular-weight phthalocyanines [35], [36]. After the second stage (CoPPC), these peaks disappear, and a broad peak at about 27° is observed. It can be interpreted as a formation of the layered structure with the inter-layer distance of about 0.3 nm. Also, in CoPPC, a peak at 5.6° emerges, that corresponds to cobalt atoms, separated from each other by ~1.5 nm in the PPC matrix [37]. The XRD pattern of the final product (CoGLC) reveals a unique structure. The (002) graphene-like peak becomes sharper, and the low-angle peaks at 5.6° and 7.9° become pronounced. These low-angle peaks provide direct evidence for a long-range ordered arrangement of cobalt atoms within the graphene-like layers.

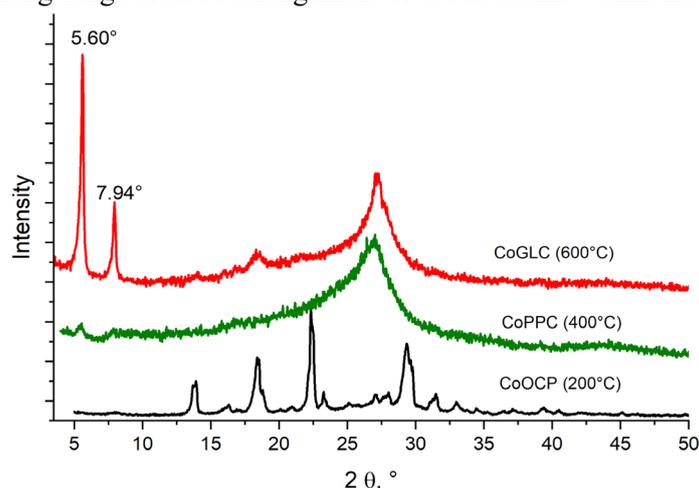

**Fig. 2.** XRD patterns of products from the three stages of the synthesis

## Infrared and Raman spectra

For CoOCP, the product of the first stage of the synthesis, both IR and Raman spectra show multiple sharp lines in the fingerprint region, and an intense line at about 2250 cm$^{-1}$, coming from terminal C≡N groups (fig. 3). After the second stage, the fingerprint lines become broader and their number decreases, which is the expected behavior upon formation of the polymeric structure [38]. Also, the intensity of the C≡N band drops significantly. For the final product, CoGLC, IR spectrum shows a continuous structure-less absorption with no pronounced signs of functional groups, which is what one should expect for graphene. In the Raman spectrum, the D, G and second-order (2D) peaks of graphene-like carbon are observed.

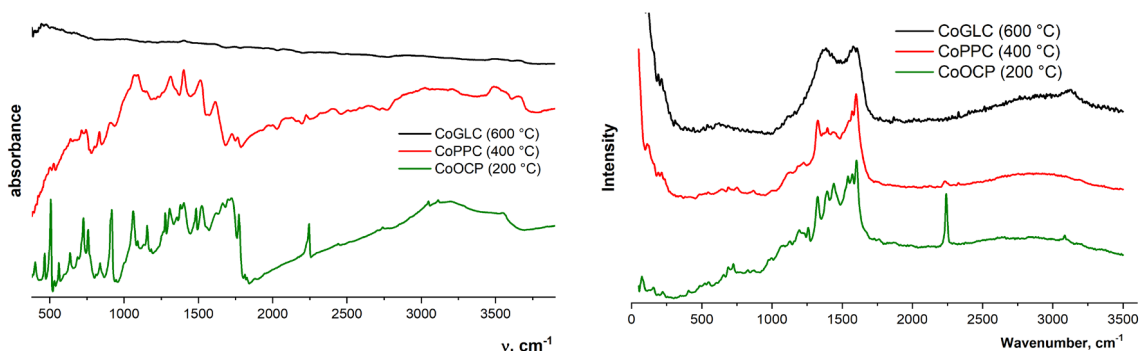

**Fig. 3.** IR (left) and Raman (right) spectra of products from the three stages of the synthesis

**Inks fabrication, TEM and conductivity**

The as-synthesized CoGLC is a black powder, which limits its practical application. To overcome this, we developed a processable conductive ink. For production of stable inks from 2D materials, a necessary step is a preliminary exfoliation [39], [40]. Herein, we employed electrochemical exfoliation [41] with subsequent ultrasonic treatment. In the electrochemical exfoliation, tetrabutylammonium cations (TBA+) intercalate between the CoGLC layers, which leads to expansion of interlayer spacing. Further delamination of CoGLC occurs during ultrasonic treatment in NMP. Next step, we replace NMP with ethylene glycol through the Pickering emulsion. The resulting functional inks are additive-free; the key role in stabilizing dispersion play interparticle van der Waals interactions [42].

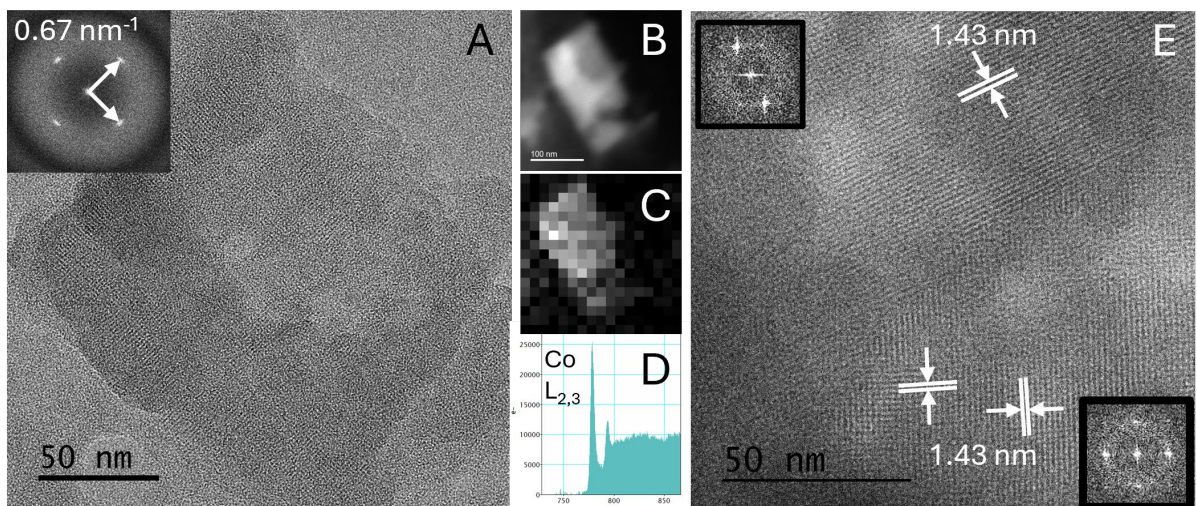

**Fig. 4**: Transmission electron microscopy of CoGLC sample and EELS mapping of cobalt. (A) Bright-field TEM image of a ~150 nm particle; the FFT power spectrum shows reflections at ≈ 0.67 nm$^{-1}$ forming a 90° angle (orthogonal lattice directions). (B) STEM-HAADF image of the particle in (A). (C) EELS map of integrated intensity under the Co $L_{2,3}$ edge. (D) Background-subtracted EELS spectrum summed over the particle. E – Thicker region showing a boundary between two differently oriented lattice domains with the same d-spacing (1.43 nm).

TEM images show that the inks consist of flake-like particles with the characteristic size of tens to hundreds nm (fig. 4). Cobalt can be easily detected in the flakes by EELS mapping. From the FFT power spectrum we may conclude that there is a distinct periodicity related to the arrangement of cobalt atoms. If analyzed together with XRD pattern and literature data, the reflection pattern can be interpreted as formed by the square lattice of cobalt atoms, that can have spacings of ~1.5 nm. Such distance can be originally formed by phthalocyanine structure if cobalt atoms are included in a diagonal pattern (1.5 nm) [37]. The similar arrangement of Co atoms is inherited by the CoGLC structure.

We demonstrate that the inks can be deposited on a dielectric substrate (SiO$_2$/Si) for electrical measurements (Fig. S4). Temperature-dependent resistivity plot can be approximately described by Arrhenius dependence, from which the charge transport activation energy can be estimated. The value of 0.15 eV was obtained.

**XPS**

XPS analysis was performed on a CoGLC ink sample, the final stage of the synthesis (Fig. 5, S5, S6). The spectra confirm the structure typical for nitrogen-doped carbon materials. In the C 1s, there is an intense signal characteristic of sp2 carbon - 285 eV, while in N 1s, the main signals correspond to nitrogen in the structure of pyridine (399.3 eV) and pyrrole (400.8 eV) fragments. Meanwhile, the C 1s and Co 2p spectra regions also contain signals specific to nitrogen-carbon complexes with cobalt: 286.5 eV – C-N, including coordinating Co nitrogen atoms and Co-N (781.4 eV – Co p3/2, 796.7 eV – Co p1/2). Signals indicating the presence of oxygen bonds are also visible in the C 1s, N 1s, and O 1s (288.02 eV, 403.2 eV, 532.3 eV, 533.5 eV). Oxygen atoms are probably present in the structure as a result of partial oxidation of the sample which could have occurred during the multi-stage synthesis, especially during the exfoliation stage. However, these signals only refer to the bonding of oxygen with carbon or nitrogen. The peaks characteristic for Co-O are absent in the O 1s region.

According to XPS data, the ratio of C:N:Co:O atoms in the ink sample is 3.94:1:0.06:0.32 (N taken as a unit). Ideally, the theoretical stoichiometric ratio for CoPPC is 2.5:1:0.125:0. Thus, there is an increased relative carbon content in our sample. This circumstance can be explained by the conditions of the carbonization process, under which the nitrogen content in 2D material decreases [43], [44], [45]. Nevertheless, the degree of nitrogen doping is still high—about 19% in terms of XPS. The reduced Co content compared to the stoichiometric value may have resulted because not all coordination centers of polyphthalocyanine were occupied by metal atoms. Also, this can be explained by the partial removal of Co atoms during the cleaning/exfoliation/carbonization stages of the ink sample preparation. However, about half of the Co remains in the structure of the 2D material.

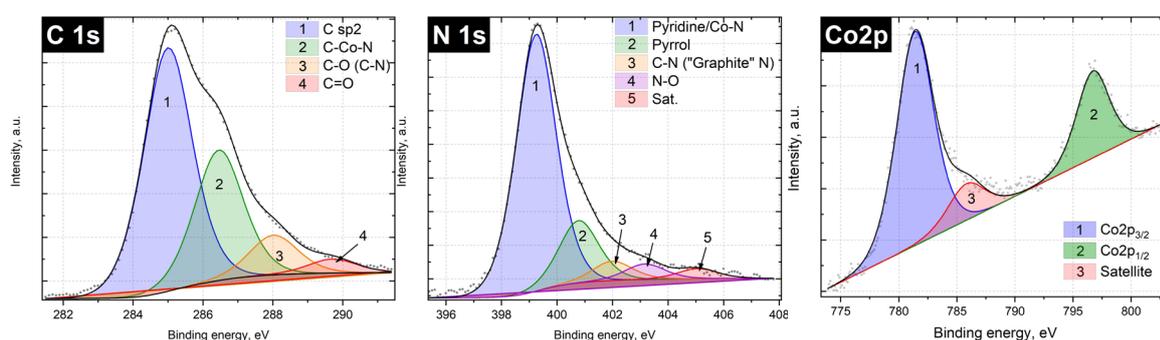

**Fig. 5**. XPS of the CoGLC film deposited from the inks: C1s (left), N1s (middle) and Co2p (right) peaks.

## Conclusions

We have designed and executed a three-stage synthesis for a novel graphene-like material, CoGLC, featuring an ordered arrangement of cobalt atoms. Both XRD and TEM studies confirm this long-range periodicity within the carbon layers. The material was successfully processed into surfactant-free, conductive inks via electrochemical exfoliation, addressing a key challenge in device integration. Temperature-dependent conductivity measurements revealed a charge transport activation energy of approximately 0.15 eV. The chemical structure and covalent bonding of Co, N, and C were verified by XPS. This work provides a pathway for creating ordered metal-carbon frameworks and demonstrates their potential for application in printable electronics and catalysis.

## Acknowledgments

The XPS data were collected using Specs spectrometer of the Shared Research Center "Surface and novel materials" of UdmFRC UB RAS.

TEM measurements were carried out at the Shared Research Facility "Electron microscopy in life sciences" at Moscow State University (Unique Equipment "Three-dimensional electron microscopy and spectroscopy").

The study was supported by the Ministry of Science and Higher Education of the Russian Federation, project № 075-15-2024-560.

Supporting information

Loading of the ampoules

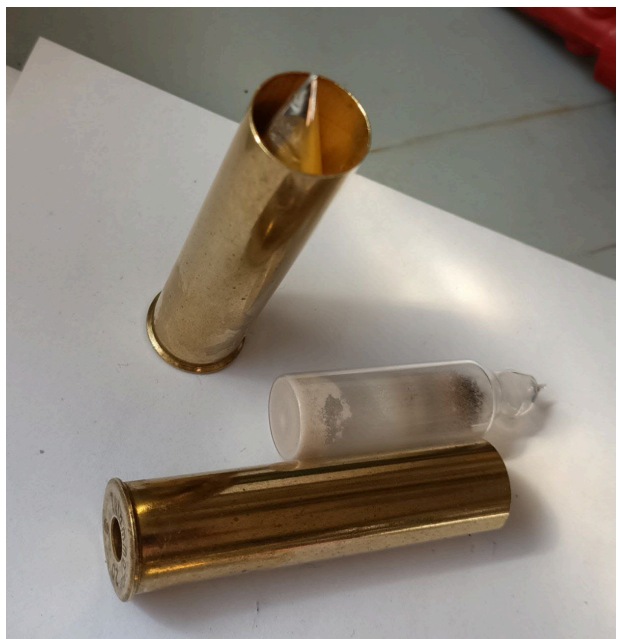

Fig. S0. Loading of the ampoules in a 12 caliber shotgun casing

# TG/DSC data

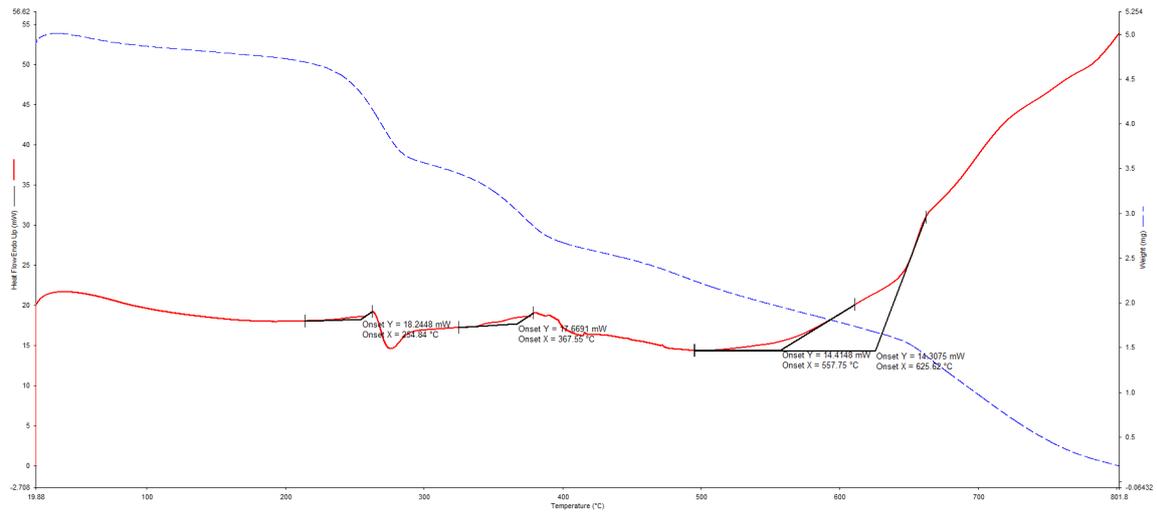

Fig. S1. Mass and heat flow plots for CoGLC

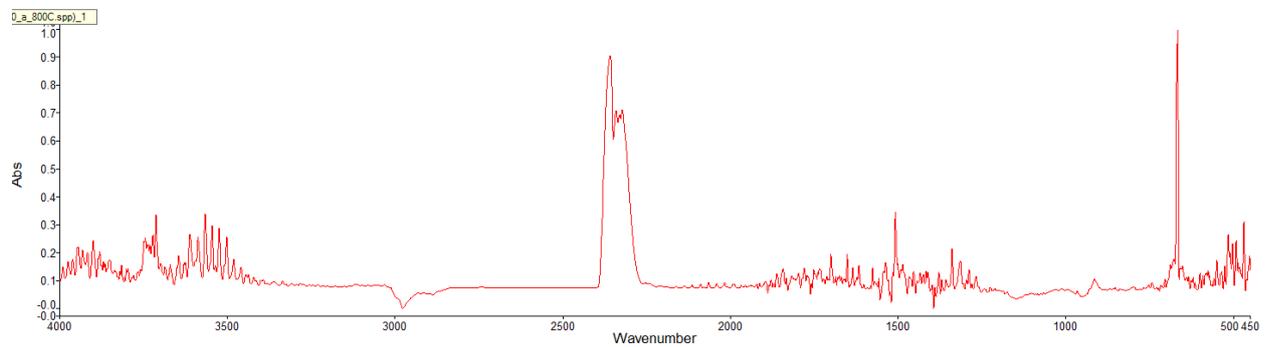

Fig. S2. IR spectra of the gas phase at ~245 °C

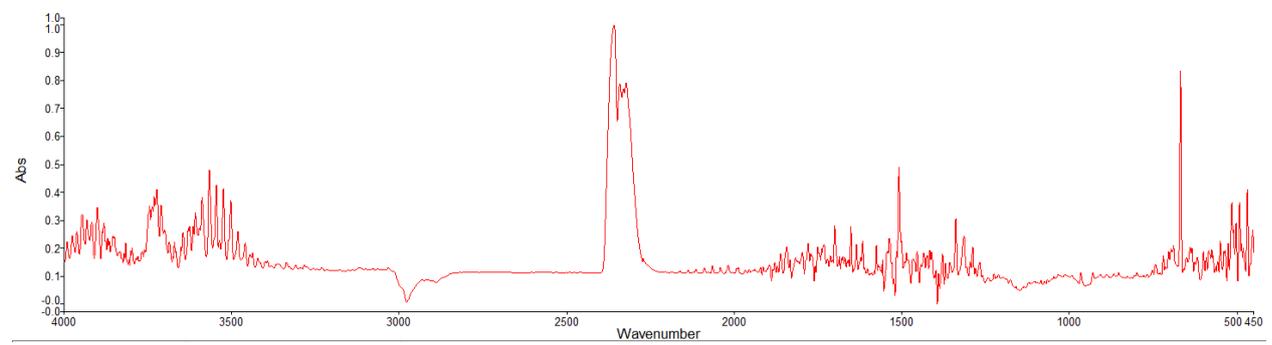

Fig. S3. IR spectra of the gas phase at ~366 °C

Electrical measurements

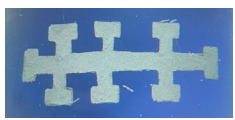

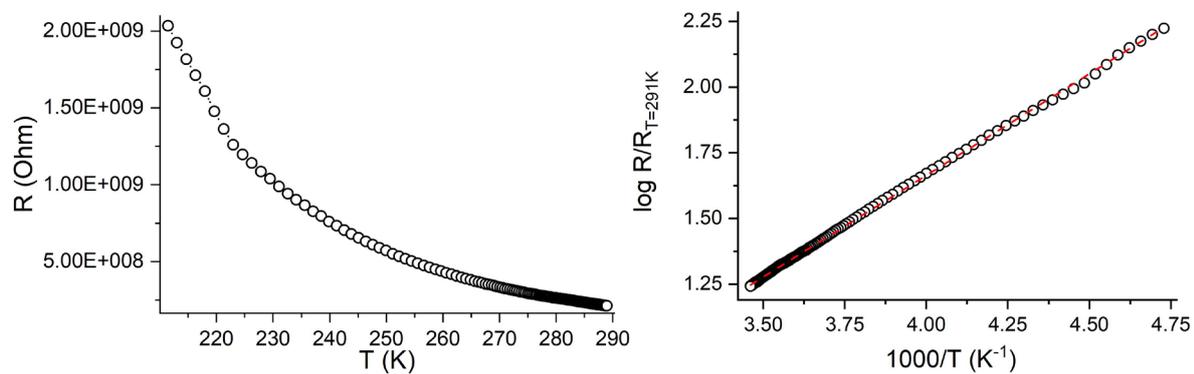

Fig. S4 Temperature-dependent resistivity measurements of the CoGLC film (bottom) and the Hall-bar pattern of CoGLC on SiO$_2$/Si substrate (top)

Additional XPS data

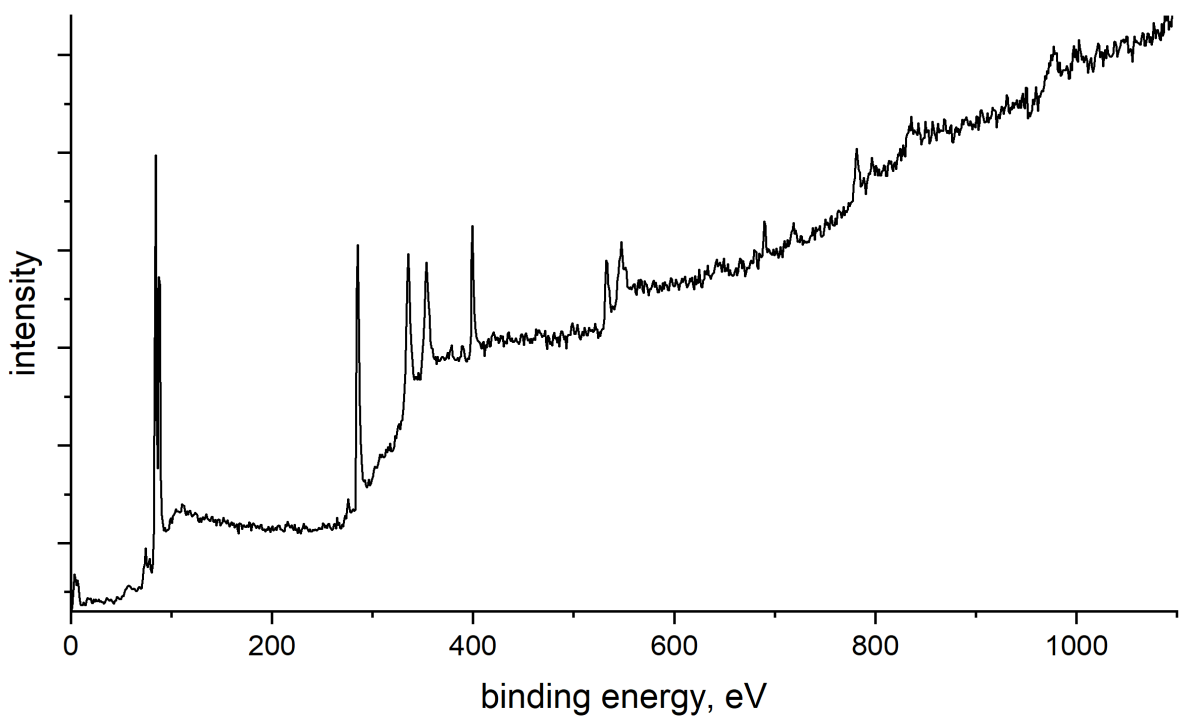

Fig. S5. Survey XPS of CoGLC

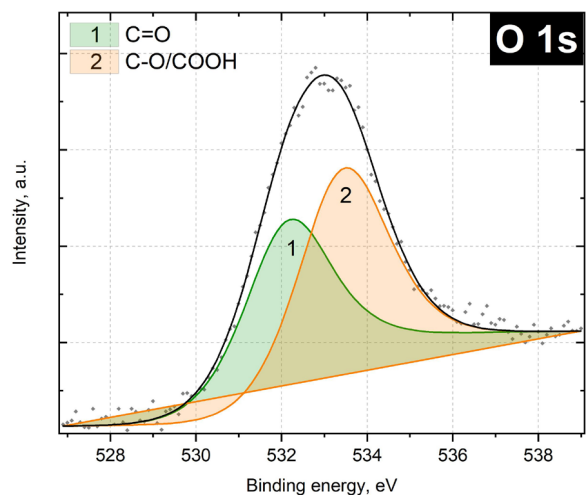

Fig. S6. O 1s peak of CoGLC